\documentclass{aa}  
\usepackage{hyperref}
\usepackage{graphicx}
\usepackage{txfonts}
\usepackage{newtxtext,newtxmath}
\usepackage{placeins}
\usepackage{lipsum}
\usepackage{algorithm}
\usepackage{algpseudocode}
\usepackage{booktabs}
\usepackage{subcaption}         
\usepackage{lscape}             
\usepackage{placeins}           
                                
\begin{document}
\nolinenumbers
\title{Radio-Interferometric Image Reconstruction with Denoising Diffusion Restoration Models}

%
%
%

   \author{M. Morales\inst{1}       
        \and E. Tolley\inst{1}\corrauth{emma.tolley@epfl.ch}   
        \and R. Poitevineau\inst{1}
        }

   \institute{Institute of Physics, Laboratory of Astrophysics, \'{E}cole Polytechnique F\'{e}d\'{e}rale de Lausanne (EPFL), 1290 Sauverny, Switzerland}

   \date{Received August 6, 2026}

 
  \abstract
   {}
   {Reconstructing images of the radio sky from incomplete Fourier information is a key challenge in radio astronomy. In this work, we  present a method for radio   interferometric image reconstruction using a data-driven prior for the radio sky based on denoising diffusion probabilistic models (DDPMs).    } 
   {We train a DDPM on radio galaxy observations from the VLA FIRST survey, then create simulated VLBA, EHT, and ALMA observations of radio galaxies. We use an unsupervised posterior sampling method called Denoising Diffusion Restoration Models (DDRM) to reconstruct the corresponding images using our DDPM as a prior. Our approach naturally incorporates the PSF of the instrument. }
   {We are able to reconstruct images with very high fidelity and demonstrate a marked improvement over  CLEAN and MS CLEAN. While DDRM naturally produces multiple samples, these are not calibrated and do not constitute reliable uncertainty estimates in the current implementation. The code for training and inference of the model is available at https://github.com/epfl-radio-astro/diffusionRI.git}
   {}


   \maketitle

\section{Introduction}
\nolinenumbers

The sparse layouts of radio interferometers result in an incomplete sampling of the sky in Fourier space. 
Aperture synthesis by radio interferometry aims to reconstruct images of the radio sky from this incomplete Fourier information. 
This ill-posed inverse problem requires  advanced image formation algorithms and strong regularization to compensate for the missing information.

Traditional imaging in radio-interferometry has relied on the CLEAN algorithm introduced in \cite{1974A&AS...15..417H}, a sparse matching pursuit algorithm. 
CLEAN has had many variants and improvements over the decades, such as Multi-scale CLEAN \citep{Cornwell_2008}, W-projection, W-stacking, w-snapshotting techniques \citep{Cornwell_2008_W_proj, 1992A&A...261..353C, Ord_2010}, which correct the non-coplanar baseline term, A-projection \citep{Bhatnagar_2008} that correct per antenna gain variation. These improvements have made CLEAN a robust algorithm that is a staple of radio-interferometric imaging.
 Despite CLEAN's success it still has certain limitations. The performance relies on user--defined parameters, the imaging lacks uncertainty quantification, and complex emission can be difficult to model with CLEAN's discrete components. 

 
Computational imaging techniques provide an alternate approach to image reconstruction. These methods model how the observations $\mathbf y$ depend on the sky model $\mathbf x$, use additional regularization terms such as sparsity priors. 
These techniques include maximum entropy
methods \citep[MEM; ][]{10.1093/mnras/163.4.369, 1974A&AS...15..383A, 1985A&A...143...77C,1986ARA&A..24..127N}, compressed sensing (CS) and sparse reconstruction methods \citep[for example,][]{10.1111/j.1365-2966.2009.14665.x,2011A&A...528A..31L,Carrillo_2012,2015A&A...576A...7D, 2023MNRAS.518..604T}, and regularized maximum likelihood imaging \citep[RML;][]{Akiyama_2019,2023PASP..135f4503Z}.

Recent machine learning approaches in radio-interferometric imaging also show promising results, learning the distribution of the radio sky rather than relying on a hand-crafted prior. \cite{2022MNRAS.509..990G} and \cite{2022MNRAS.514.2614C} developed neural networks which learn the mapping between the dirty image and the true radio sky. \cite{2022A&A...664A.134S} redefined the deconvolution problem as inpainting in Fourier space, using a residual neural network~\citep{7780459}. \cite{2024ApJS..273....3A} developed a series of Deep Neural Networks (DNNs) and formulate reconstruction as a series of images, with each DNN taking the previous iteration’s image estimate
and associated data residual as inputs. \cite{2024A&A...683A.105D} and \cite{Wang2023ConditionalDDPM} developed conditional Denoising Diffusion Probabilistic Models~\citep[DDPM;][]{ho2020denoisingdiffusionprobabilisticmodels} to deconvolve dirty images. 

However, all of these machine learning methods rely on learning a specific antenna configuration. These networks must be retrained to apply to different telescopes, or even to use for observations with different observing times or pointing directions. 

In this paper we present a deep learning method for interferometric image reconstruction that does not need to be trained on any specific antenna configuration, as it naturally incorporates the physics of the measurement process.  First, we train a DDPM which learns the morphology of radio galaxies from VLA FIRST~\citep{becker1995first} survey which serves as a data-driven prior. Then we implement image reconstruction method using a chain of conditional reconstruction steps called Denoising Diffusion Restoration \citep[DDRM;][]{kawar2022denoisingdiffusionrestorationmodels} to obtain samples consistent both with the noisy observation and with the distribution of training data. We evaluate our results using a simplified reconstruction pipeline of the VLBA array layout, and compare the performance to CLEAN,   MS CLEAN, IUWT compressed sensing, uSARA, and AIRI.

 The paper is organized as follows. Section~\ref{sec:methods} covers our methodology, including a formulation of image reconstruction as an inverse problem, an introduction to DDPMs and DDRM sampling, an overview of the data used for training and validation, and our process for creating our mock observations. Results are presented in~\ref{sec:results}, including comparisons to other deconvolution methods, tests with variable noise levels, and imaging out-of-domain data.

\begin{figure*}
  \centering
  \includegraphics[width=.9\linewidth]{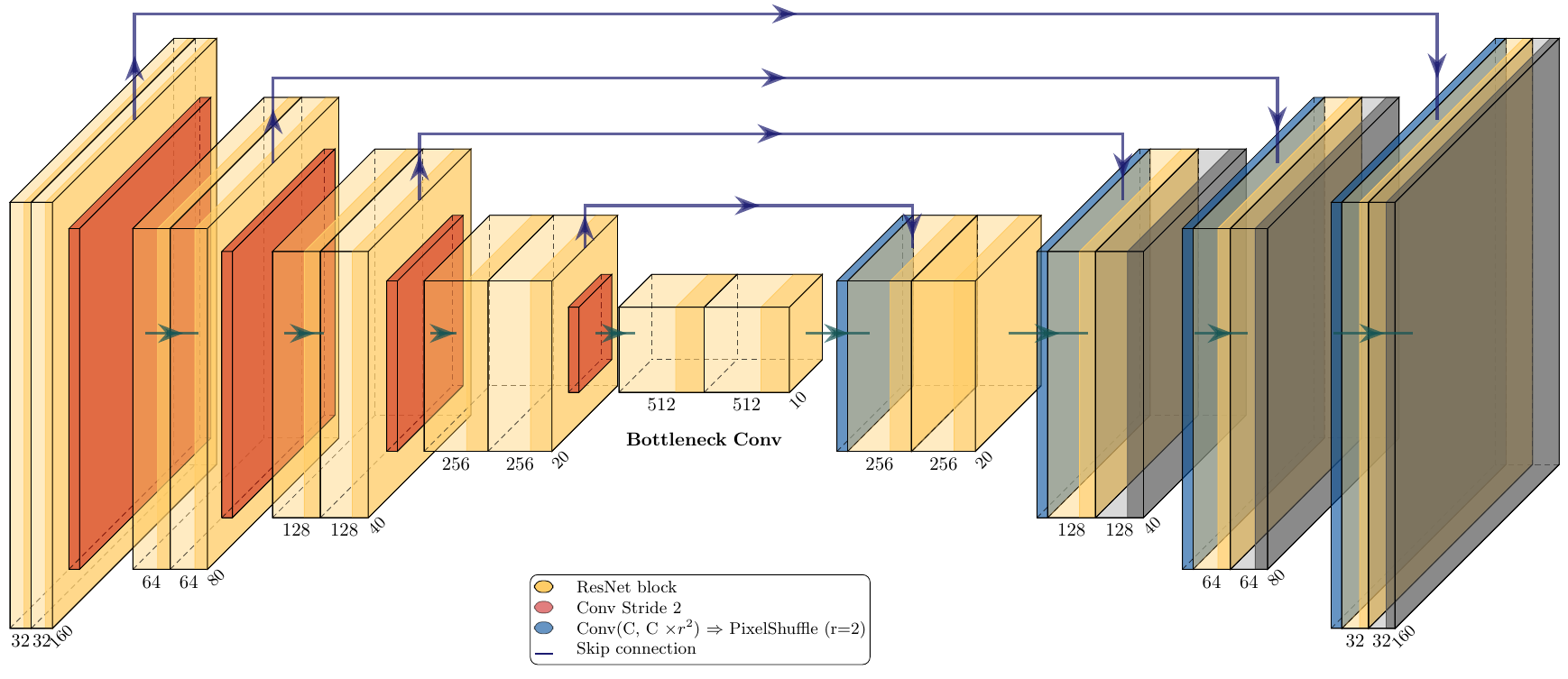}
  \caption{Neural Network Architecture used for the DDPM network.}
  \label{fig:network_architecture}
\end{figure*}

\section{Methods}
\label{sec:methods}
\subsection{Aperture synthesis in radio interferometry}
\label{sec:ri}

A radio interferometer measures the spatial coherence function of the electric fields measured at antenna positions $\vec r_p$ and $\vec r_q$~\citep{Cornwell_2008_W_proj}:
\begin{equation}
    V_{pq} = \langle E(\vec r_p,t) E(\vec r_q,t)^*\rangle_t,
\end{equation}
These measurements, usually called visibilities, are often expressed as $V(u,v,w)$, where $u, v, w$ are components of the vector between the two interferometer elements expressed in units of wavelength of the radiation.
The visibilities can be related to the sky brightness distribution $I(l,m)$ via the van Cittert-Zernike theorem~\citep{2009OExpr..17.1746O}:
\begin{equation}
    V(u,v,w) = \iint I(l,m) e^{-2 \pi i [ul + vm + wn]}\frac{dl \, dm }{n+1} + z(u,v,w),
\end{equation}
where $n := \sqrt{1-l^2-m^2}-1$ and $z$ are Gaussian-distributed uncorrelated noise terms. In the small-field approximation where $n \approx 0$, we can rewrite this in matrix notation~\citep{10.1093/mnras/stx113} as:
\begin{equation}
    \mathbf{y = S \mathcal{F} x + z}, \label{eq:fwdmodel0},
\end{equation}
where the visibilities $\mathbf{y} \in \mathbb{C}^m$ for $m$ antenna pairs, $\mathbf{x} \in \mathbb{R}^n $ is the flattened radio sky image,  $\mathcal{F} \in \mathbb{C}^{n \times n}$ is the 2D Fourier transform matrix, $\mathbf{S} \in  \mathbb{C}^{m \times n}$ is the sampling matrix due to the finite number of antennas, and $\mathbf{z} \in \mathbb{R}^m $ are uncorrelated noise terms. We note that the sampling matrix $\mathbf{S}$ depends on the interferometer layout and observation configuration. 

The CLEAN image reconstruction algorithms start by reconstructing an initial estimate of $\mathbf{x}$ through back-projection, called the ``dirty image'':
\begin{equation}
    \mathbf{x}_D = \mathcal{F}^{-1}\mathbf{y} = \mathcal{F}^{-1}(\mathbf{S})*\mathbf{x} + \mathcal{F}^{-1}\mathbf{z}, \label{eq:dirty_image}
\end{equation}
and the ``dirty beam'' $\mathcal{F}^{-1}(\mathbf{S})$ is iteratively removed from the dirty image $\mathbf{x}_D$ by
populating a model of the radio sky with discrete components.
%
Hogbom CLEAN~\citep{1974A&AS...15..417H} uses point sources as its components, whereas Multi-Scale (MS) CLEAN~\citep{Cornwell_2008, 2008AJ....136.2897R} uses point and extended sources modeled with a tapered quadratic function. Both variations of CLEAN are \emph{matching pursuit} algorithms that approximately maximize the posterior with an implicit sparsity prior~\cite{529037}.  The compressed sensing algorithm MORESANE~\citep{2015A&A...576A...7D} adds components across multiple scales using the Isotropic Undecimated Wavelet Transform \citep[IUWT][]{2007ITIP...16..297S}, which is also implemented in the WSCLEAN library as IUWT CS.

Imaging can also be constructed as a minimization problem:
\begin{equation}
    \min_\mathbf{x} \frac{1}{2} ||\mathbf S \mathcal{F} \mathbf x - \mathbf y||^2 + \lambda ~r(\mathbf x)
\end{equation}
where $\lambda >0$ is a regularization parameter an $r$ is a prior on $\mathbf x$.
The uSARA~\citep{Carrillo_2012, 9053284} algorithm reconstructs a radio image by solving the above optimization problem by enforcing sparsity with a multi-scale wavelet dictionary, while the AIRI~\citep{2023MNRAS.518..604T}  algorithm uses a learned denoiser instead of the explicit sparsity prior.



    
%

\subsection{Denoising Diffusion Probabilistic Models }
\label{sec:ddpm}

For more details about DDPM, its theoretical foundation, capabilities, and training algorithms we refer the reader to \cite{ho2020denoisingdiffusionprobabilisticmodels}.
The DDPM process is defined by two Markov chains, a forward (diffusion) and backwards (denoising) process. The forward process ${q(\mathbf{x}_{1:T} | \mathbf{x}_0)}$ 
gradually adds Gaussian noise to the original data $\mathbf{x}_0$ over $T$ steps, producing a sequence of increasingly noisy samples $\mathbf{x}_1,\dots,\mathbf{x}_T$:
\begin{equation}
    q(\mathbf{x}_{1:T} | \mathbf{x}_0) 
    := \prod_{t = 1}^{T} q(\mathbf{x}_t | \mathbf{x}_{t-1}) .\label{eq:ddpm_forward}
\end{equation}
Each transition $q(\mathbf{x}_t|\mathbf{x}_{t-1})$ 
adds noise which takes the form of a Gaussian distribution with decreasing signal-to-noise ratio determined by the schedule $\beta_t$:
\begin{equation}
    q(\mathbf{x}_t | \mathbf{x}_{t-1})
    := \mathcal{N}(\mathbf{x}_t ; \sqrt{1 - \beta_t}\mathbf{x}_{t-1}, \beta_t \mathbf{I}).
    \label{eq:ddpm_forward_step}
\end{equation}
%
The reverse process $p_\theta(\mathbf{x}_{0:T})$ 
is learned by a neural network to approximate the reverse of the forward diffusion:
\begin{equation}
    p_\theta(\mathbf{x}_{0:T}) := p(\mathbf{x}_T) \prod_{t=1}^T p_\theta(\mathbf{x}_{t} | \mathbf{x}_{t+1}), \label{eq:ddpm_reverse}
\end{equation}
where $\theta$ refers to the parameters of the neural network.
This Markov chain starts from a sample $\mathbf{x}_T \sim \mathcal{N}(0, I)$ and attempts to remove a small amount of Gaussian noise at each step. The step function is modeled with:
\begin{equation}
    p_\theta(\mathbf{x}_{t} | \mathbf{x}_{t+1}) := \mathcal{N}(\mathbf{x}_{t}; f_\theta(\mathbf{x}_{t+1}, t), \sigma_t^2 \mathbf{I}) \label{eq:ddpm_reverse_step}
\end{equation}
where the mean function $f_\theta(\mathbf{x}_{t+1}, t)$ is the output of the neural network and $\sigma_t^2$ is the accumulated variance noise at each timestep $t$.
The operation is repeated $T$ times to recover the denoised sample $\mathbf{\hat{x}}_0$.


\subsubsection{Network Architecture}

We implement a DDPM using the U-Net architecture \citep{2015arXiv150504597R}, a multi-scale convolutional neural network with skip-connections to preserve fine detail. The network has two parts, an encoder that progressively compresses the image to higher-level features, and a decoder that progressively reconstructs the image from these features. 
Each encoder and decoder block is made up of two Resnet \citep{2016cvpr.confE...1H} blocks with FiLM layers \citep{2017arXiv170907871P} after each LayerNorm~\citep{2016arXiv160706450L}. The network has a total of 40 million free parameters. A diagram of the network architecture is shown in Figure~\ref{fig:network_architecture}.

\subsection{Denoising Diffusion Restoration models for radio-interferometry}
\label{sec:ddrm}
 
\cite{kawar2022denoisingdiffusionrestorationmodels} introduced the diffusion restoration model (DDRM), which defines a
Markov chain conditioned on observations $y$ which performs approximate posterior
sampling for $p(x_0|y)$, where the prior on the
true image $p(x_0)$ is implicitly given by a pretrained diffusion model.
DDRM samples from a constrained generative process:
\begin{equation}
    p_\theta(\mathbf{x}_{0:T} | \mathbf{y}) := p(\mathbf{x}_T|\mathbf{y}) \prod_{t=1}^T p_\theta(\mathbf{x}_{t-1} | \mathbf{x}_t,\mathbf{y}).
\end{equation}
The measurements $\mathbf{y}$ are obtained from the true image $\mathbf{x}$ from a linear forward model:
\begin{equation}
    \mathbf{y} = \mathbf{H}\mathbf{x} + \mathbf{z}, \label{eq:fwdmodel}
\end{equation}
where $\mathbf{H}$  is the linear transformation operator and $\mathbf{z}$ is zero-mean Gaussian noise with variance $\sigma_y^2$. 
Their solution is expressed in the singular
space of the sampling operator and therefore starts with computing its singular value decomposition (SVD):
\begin{equation}
    \mathbf{H} = \mathbf{U\Sigma V}^\top,
\end{equation}
then applying the transformations $ \mathbf{\bar y := \Sigma^+U^\top y}$ (where $\mathbf{\Sigma^+}$ is a Moore–Penrose pseudo-inverse) and $\mathbf{\bar x := V^\top x}$. If $f_\theta(\mathbf{x}_{t+1}, t)$ is the prediction made by the DDPM model at timestep $t+1$, we define $\mathbf{\bar x}^{}_{\theta,t} := \mathbf{V^\top} f_\theta(\mathbf{x}_{t+1}, t)$. The sampling procedure then considers separately components according to the singular values of the transformation operator.

For components with zero singular value ($s_i=0$), we have  no information from the measurements $\mathbf{y}$ and sample according to the prior only:
\begin{equation}
p_{\theta}\!\left(\mathbf{\bar{x}}^{(i)}_t \mid \mathbf{x}_{t+1}, \mathbf{y}\right)
= \mathcal{N}\!\left(
\bar{x}^{(i)}_{\theta,t}
+ \sqrt{1-\eta^2}\,\sigma_t
\dfrac{\mathbf{\bar{x}}^{(i)}_{t+1}-\mathbf{\bar{x}}^{(i)}_{\theta,t}}{\sigma_{t+1}},
\;\eta^2\sigma_t^2
\right).
\end{equation}
When measurement is uncertain ($\sigma_t < \sigma_y/s_i$), the update is guided by the measurement and weighted based on the uncertainty:
\begin{equation} \label{eq:sampleuncertain}
p_{\theta}\!\left(\mathbf{\bar{x}}^{(i)}_t \mid \mathbf{x}_{t+1}, \mathbf{y}\right)
= \mathcal{N}\!\left(
\mathbf{\bar{x}}^{(i)}_{\theta,t}
+ \sqrt{1-\eta^2}\,\sigma_t
\dfrac{\mathbf{\bar{y}}^{(i)}-\mathbf{\bar{x}}^{(i)}_{\theta,t}}{\sigma_y/s_i},
\;\eta^2\sigma_t^2
\right).
\end{equation}
When the measurement is certain compared to the prior's variance schedule ($\sigma_t \ge \sigma_y/s_i$), the variance shrinks accordingly:
\begin{equation}
p_{\theta}\!\left(\mathbf{\bar{x}}^{(i)}_t \mid \mathbf{x}_{t+1}, \mathbf{y}\right)
= \mathcal{N}\!\left(
(1-\eta_b)\mathbf{\bar{x}}^{(i)}_{\theta,t} + \eta_b \mathbf{\bar{y}}^{(i)},
\;\sigma_t^2 - \dfrac{\sigma_y^2}{s_i^2}\eta_b^2
\right)
\end{equation}
In all three cases $\eta$ and $\eta_b$ are  hyperparameters that respectively control the amount of stochastic noise injected during sampling and the strength with which the measurement is enforced in the reverse process. 
 When setting $\eta_b = 1.0$, the sampling procedure will exclusively update using the observation $\mathbf{y}$ if the measurement noise is smaller than the variance schedule weighted by the singular value. An optimal choice of $\eta$ depends on the number of sampling steps used and the desired tradeoff between reconstruction accuracy and uncertainty.
 In this work we use $\eta = 0.65$ and $\eta_b = 1.0$, which we found had a good tradeoff between reconstruction quality and uncertainty for $1000$ sampling steps, shown in Appendix~\ref{sec:eta_tuning}.

To adapt the forward model in equation~\ref{eq:fwdmodel} and corresponding SVD for aperture synthesis, we define the transformation operator $\mathbf{H = S}  \mathcal{F}$.  We first find a memory-efficient SVD of $\mathbf{S}$ following the method presented in \cite{kawar2022denoisingdiffusionrestorationmodels}:
\begin{equation}
\mathbf{S = I K P },
\end{equation}
where $\mathbf{P}$ is an appropriate permutation matrix, $\mathbf{K}$ is a rectangular diagonal matrix of size $m \times n$ with the gridded visibility weights in its main diagonal, and $\mathbf{I}$ is the identity matrix. Because $\mathcal{F}$ is unitary, we can easily define the complex SVD:
 
$\mathbf{H} = \mathbf{U\Sigma V}^*$,
where $\mathbf{U = I}$, $\mathbf{\Sigma = K}$, and  $\mathbf{V^* = P\mathcal{F}}$, with $\mathbf{V = \mathcal{F}^{-1}P^\top}$. This accurately represents the radio astronomy measurement equation in the small-field approximation, but does not include the effect of $w-$terms or $A-$terms.   Instead of calculating the Fourier matrices directly, we implement them with the pytorch implementation of the fast Fourier transform FFT2.

The general computational complexity of our method is $\mathcal{O}(K \cdot (C(N) + S(N)))$, where K is the number of sampling steps, $C(N)$ and $S(N)$ are the complexity of the neural network and the measurement consistency step, with respect to the number of pixels $N$.
In our neural network, our complexity is $C(N) = \mathcal{O}(N)$ and our data consistency step complexity is dominated by the FFT2 algorithm, hence $S(N) = \mathcal{O}(N \cdot log(N))$. Therefore the total complexity of  our method is $\mathcal{O}(K \cdot N \cdot log(N))$.


\subsection{Data \& Training}

\begin{figure}
    \centering
    \includegraphics[width=0.9\linewidth]{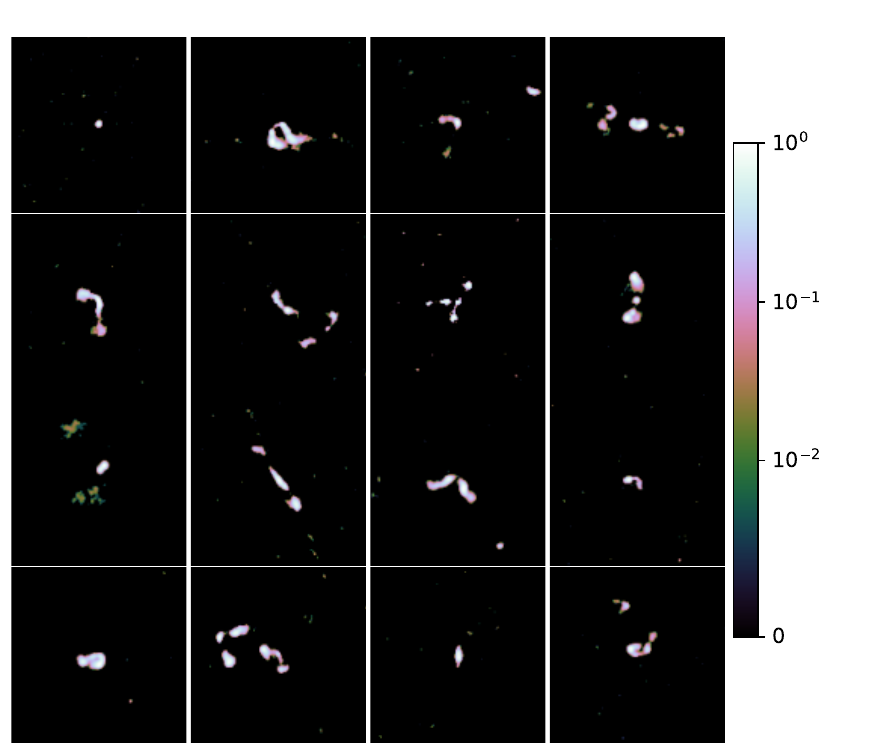}
    \caption{A selection of radio galaxies from the test dataset.}
    \label{fig:data}
\end{figure}

Before using DDRM, we first need to train our DDPM model which serves as the data-driven prior. For this we used two different overlapping radio galaxy datasets.
The first is a collection and combination of several catalogues using the VLA FIRST (Faint Images of the Radio Sky at Twenty-Centimeters) survey~\citep{becker1995first}, curated by~\cite{2023Griese}. It contains galaxies identified as FRI, FRII, compact, or bent sources. We also use 20k FIRST radio galaxies from the Radio Galaxy Zoo DR1~\citep{2025MNRAS.536.3488W}. These datasets contain images of radio galaxies of size $150\times150$ pixels and $300\times300$ pixels, respectively, {  and are flux-normalized to the range [0,1]. We crop the larger images to $150\times150$ for consistency between the two datasets.} This sample of radio galaxies comprise the set of our true images $\{\mathbf{x}\}$. We use 200 images for validation data, 200 for test data, and 21,758 images for training following the validation and test split of \cite{2023Griese}. 
A selection of radio galaxies from the test dataset are shown in Figure~\ref{fig:data}.

For training the model, we use data augmentation, including random shifts with a maximum distance of 20 pixels, random rotations {  of multiples of 90 degrees}, vertical and horizontal flips, as well as multiplying each pixel in the image by a single randomly chosen value between $0.8$ and $1.2$, then clipped back to $[0,1] $   before being mapped to [-1,1] for DDPM training or DDRM sampling.

We train our 40-million parameter DDPM neural network for $300'000$ optimizer steps, using the DDPM training algorithm described in \cite{ho2020denoisingdiffusionprobabilisticmodels}. We use PyTorch's AdamW \citep{2017arXiv171105101L} with $\beta_1 = 0.9$ and $\beta_2 = 0.999$ and  weight decay $1\times10^{-2}$. The learning rate follows a linear schedule from optimizer step $0$ to $15'000$, going from $2\times 10^{-7}$ to $2\times10^{-4}$, then we use a cosine annealing schedule \citep{2016arXiv160803983L}. The model was compiled using \texttt{torch.compile} with the \texttt{inductor} backend in \texttt{max-autotune} mode with static shapes (\texttt{dynamic=False)}. This gives us a model that is able to generate images from our training dataset unconditionally, starting from Gaussian noise.

As an additional test image, we use an image of the radio galaxy {  3c353} from~\cite{2023MNRAS.518..604T}. We create several $150 \times 150 $ pixel cutouts of this radio galaxy through cropping and resampling.

\label{sec:data}

\subsection{Experimental Setup}
\label{sec:setup}

\begin{figure}
  \centering
  \includegraphics[width=1.0\linewidth]{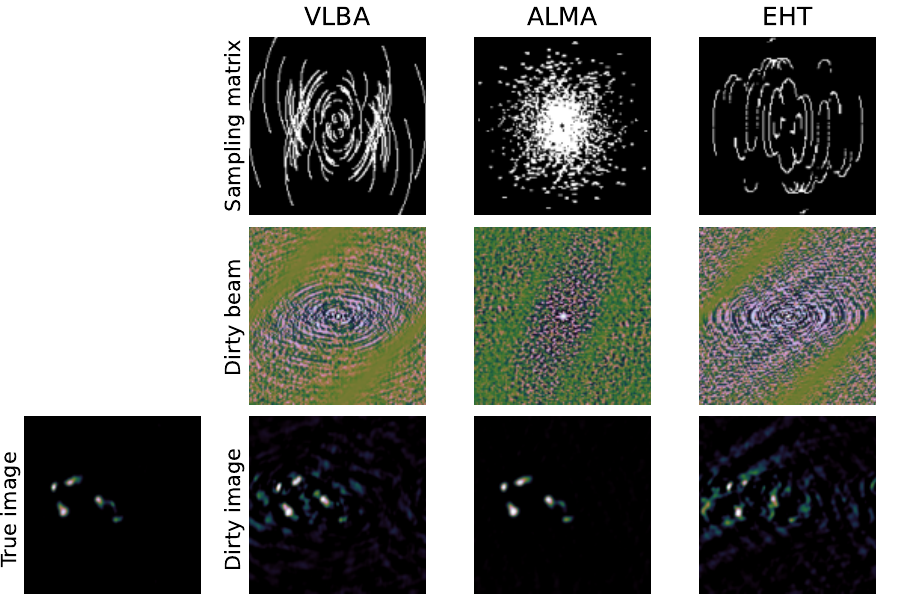}
  \caption{Diagram of the sampling matrix $\mathbf{S}$, dirty beam $\mathcal{F}^{-1}\mathbf{S}$, true image$\mathbf x$, dirty image $\mathbf x_D = \mathcal{F}^{-1}(\mathbf{S})*\mathbf x $ for the three telescopes considered in this work.}
  \label{fig:psf}
\end{figure}

To define a realistic sampling matrix $\mathbf{S}$ we use three different telescope configurations:
\begin{enumerate}
    \item { \textbf{VLBA}: We use a simulation of the  Very Long Baseline Array (VLBA)} for 60 time steps, representing approximately $2$ hours of observation time, implemented with the radio interferometer observation simulation tool from the RadioNets library~\citep{2022A&A...664A.134S}. { The baselines are gridded using a simple count-in-cell method.}
    \item \textbf{EHT}: We use a sampling matrix representing the ($u,v$) coverage the Event Horizon Telescope \citep[EHT;][]{2019ApJ...875L...2E} telescope from \cite{Wang2023ConditionalDDPM}, resized to match our $150\times150$ pixel input data {  using nearest-neighbor resampling}.
        \item \textbf{ALMA}: We use a sampling matrix representing the ($u,v$) coverage the Atacama Large Millimeter Array (ALMA)  from \cite{2023A&A...674A.161T} resized to match our $150\times150$ pixel input data  {  using nearest-neighbor resampling}.
\end{enumerate}
All three configurations correspond to a super-resolution factor of approximately 1. We construct the visibilities $\mathbf{y}$ and dirty images following Eq.~\ref{eq:dirty_image}. The sampling matrix $\mathbf{S}$, dirty beam $\mathcal{F}^{-1}\mathbf{S}$, and example dirty image $\mathbf x_D = \mathcal{F}^{-1}(\mathbf{S})*\mathbf x $ is shown for VLBA, ALMA, and EHT in Figure~\ref{fig:psf}.    Most results in this work show image reconstruction results for an $\mathbf S$ that corresponds uniform UV weights, but show additional tests with natural weights in Appendices~\ref{sec:app:weights} and ~\ref{sec:app:ood}.

We also add $\sigma_y$ noise to the dirty images. As the input sky image $\mathbf{x}$ is always scaled from [0,1], the noise added is defined as a fraction of the maximum flux of the image. 

Once we have our measurement $\mathbf{y} = \mathcal F\mathbf x_D$, we then run the DDRM sampling algorithm using our trained DDPM model following the methodology described in Section~\ref{sec:ddrm} to reconstruct a clean image of the radio sky. 

We note that because we train our DDPM to reconstruct images in the range [-1,1], the input observations need to be rescaled to be consistent with this range. If the PSF $\mathcal{F}^{-1} \mathbf S$ is normalized to $\max(\mathcal{F}^{-1} \mathbf S) = 1$, then $\max(\mathbf x_D)$ is a reasonable approximation for $\max(\mathbf x) $ and we can use it as our scaling factor. The input dirty image needs to be rescaled as:
\begin{equation}
    \mathbf x_D’ =  2\frac{\mathbf x_D}{\max( \mathbf x_D)}- 1.
\end{equation}
The observation $\mathbf y$ can then be constructed from Fourier transform of $\mathbf x_D'$: $\mathbf y' = \mathcal F \mathbf x_D'$. DDRM sampling is run on $\mathbf y$ to obtain the reconstruction $\mathbf{ \hat x}'$. Then one can recover the original units with:
\begin{equation}
    \hat x = \max( \mathbf x_D)\frac{\mathbf{ \hat x}'+1 }{2}.
\end{equation}

  \begin{figure*}
\centering
  \includegraphics[width=.99\linewidth]{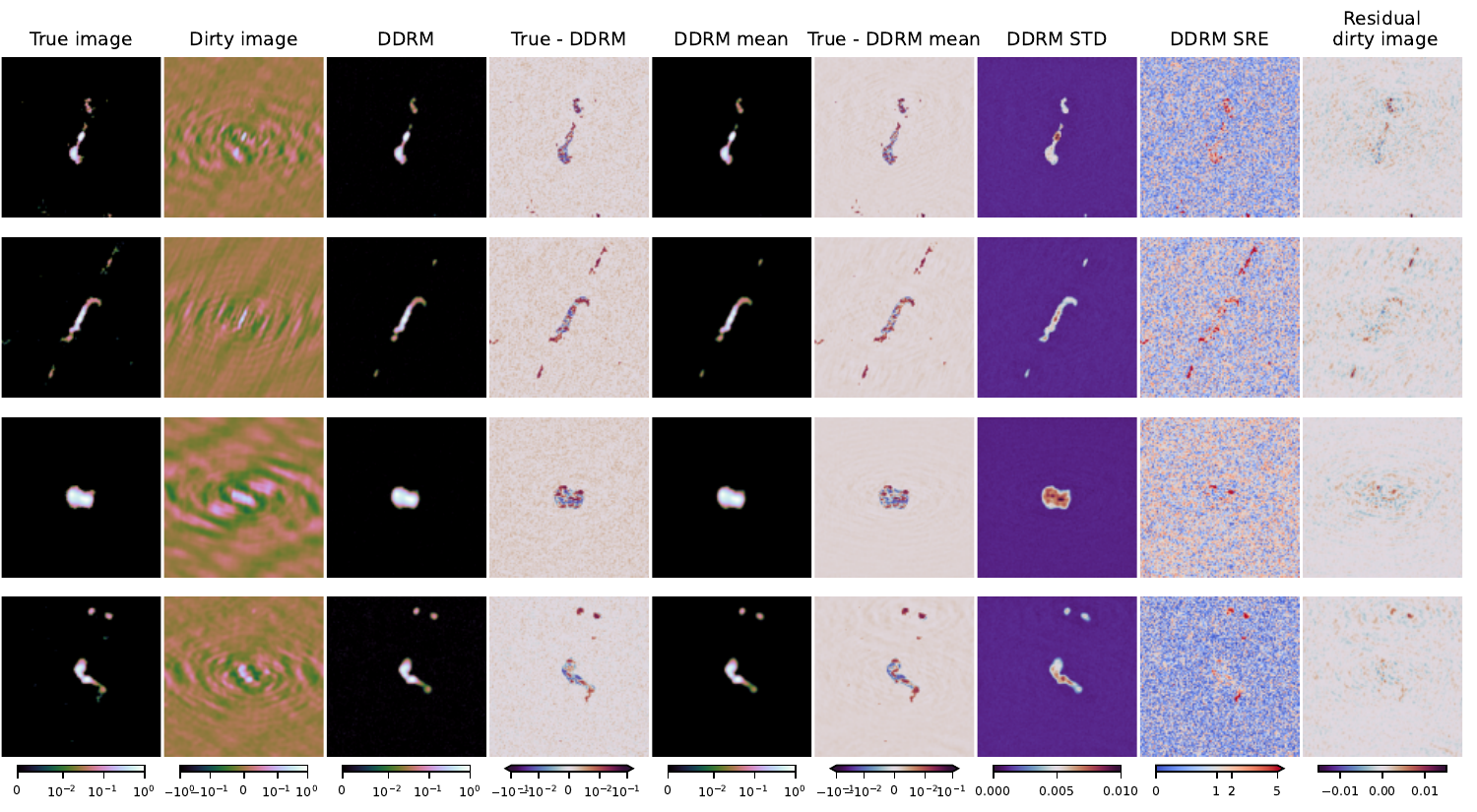}
  \caption{ DDRM reconstruction results using 1000 sampling steps for four radio galaxies from the test data set using a simulated VLBA observation, with no additional noise added. Columns from left to right are: the true image $\mathbf{x}$, the dirty image $\mathbf{x}_D$, the DDRM restored image, the residual between the DDRM image and the true image, the mean DDRM image, the residual between the mean DDRM image and the true image, the per-pixel standard deviation map, and the SRE. The plots in the last four columns are calculated across 128 DDRM reconstructions. The true and dirty images are scaled to the range [0,1], and the flux does not correspond to physical units. All log scale colors revert to linear scale in the range [0,0.01].
  }
  \label{fig:ddrmreco}
\end{figure*}

\subsection{Metrics}
We use three metrics for evaluating the quality of the results. The first is Mean Squared Error (MSE), defined as: 
\begin{equation}
    \mathrm{MSE} =  \frac{1}{NM}\sum_{i=1}^N\sum_{j=1}^M  \Big(\mathbf{x}^{(j)} - \mathbf {\hat{x}}_i^{(j)}\Big)^2,
\end{equation}
where $\mathbf x$ is our true image, $\mathbf{\hat{x}}_i$ is our $i$th prediction, and $j$ indexes the $M$ pixels of the image.
We also report the peak signal-to-noise ratio (PSNR) for comparison to other methods, defined as:
\begin{equation}
    \mathrm{PSNR} =  10 \times \log_{10} \left(\frac{\mathrm{ MAX}^2}{\mathrm{MSE}} \right),
\end{equation}
where $\mathrm MAX = 1$ for all images in our dataset. The PSNR is computed per image and reported as a mean PSNR per image

Finally, we evaluate the standardised reconstruction error (SRE), by measuring the reconstruction error relative to the reconstruction variability:
\begin{equation}
    \mathrm{SRE} = \frac{1}{NM}\sum_{i=1}^N\sum_{j=1}^M \frac{(\mathbf{\hat{x}}_{i}^{(j)}- \mathbf x^{(j)})^2}{ (\sigma^{(j)})^2}
\end{equation}
where $\sigma^{(j)}$ is the per-pixel standard deviation map:
\begin{equation}
    \sigma^{(j)} = \sqrt{
\mathbb{E}\!\left[
\bigl(\mathbf{\hat x}^{(j)}-\mathbb{E}[\mathbf{\hat x}^{(j)}]\bigr)^2
\right]
}.
\end{equation}
 Like MSE, the SRE penalizes large errors in solution, but also takes into account if the predicted solution is within the empirical reconstruction variability. 

\section{Results}
\label{sec:results}

After training our DDPM on the training data, we evaluate the DDRM reconstruction on the reserved test data.
  The results of our image reconstruction with DDRM  are shown in Figure~\ref{fig:ddrmreco}, for our simulated VLBA observation with no added noise. DDRM is able to reconstruct the central morphology of the sources with high accuracy and no remaining dirty beam artifacts.  

DDRM will occasionally miss low-flux components of the images in the reconstruction. However, these low-flux components often appear in the standard deviation map of the DDRM reconstructions, indicating that DDRM is occasionally able to recover these components.

We report the average MSE and PSNR for the VLBA, EHT, and ALMA array configurations in Table~\ref{tab:mse_calibration_vs_steps}. We find that the reconstruction accuracy is best for ALMA and worst for the EHT, corresponding with the amount of uv coverage for each observation as shown in Figure~\ref{fig:psf}.

  \begin{figure*}
\centering
  \includegraphics[width=.9\linewidth]{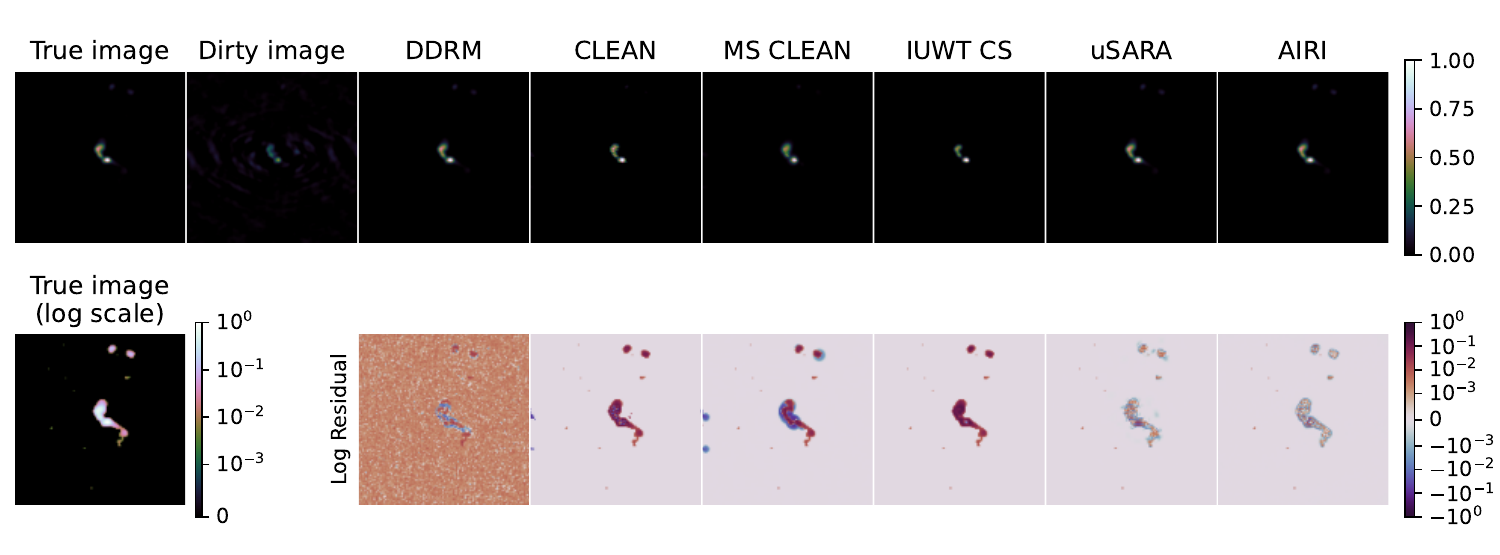}
  \caption{ Reconstruction results for a radio galaxy from the test data set using a simulated VLBA observation, with no additional noise added. Columns from left to right are: the true image $\mathbf{x}$, the dirty image $\mathbf{x}_D$, the DDRM restored image and its residuals, and then the CLEAN, multi-scale CLEAN, and IUWT compressed sensing, { AIRI, and uSARA reconstructions.} The true and dirty images are scaled to the range [0,1], and the flux does not correspond to physical units. All log scale colors revert to linear scale in the range [-0.001,0.001].
  }
  \label{fig:all}
\end{figure*}

\subsection{Sampling steps \& reconstruction quality}

\begin{table}
\centering
\begin{tabular}{cccccc}
\specialrule{.3em}{0.2em}{0.2em} 
\textbf{\emph K} & \textbf{MSE} & \textbf{PSNR} & \textbf{SRE} & $t_\mathrm{sampling}$ (s)  \\
\specialrule{.2em}{0.2em}{0.2em} 
\multicolumn{5}{c}{\textbf{VLBA array configuration}} \\
\specialrule{.05em}{0.2em}{0.2em} 
10   & $2.6 \times 10^{-4}$ & 37.0 & 1.3 & 0.44\\
50   & $6.1 \times 10^{-5}$ & 44.3 & 7.9 &  2.21  \\
100  & $2.7 \times 10^{-5}$ & 47.5 & 4.3 & 4.41 \\
500  & $6.2 \times 10^{-6}$ & 53.7 & 0.3 & 22.03 \\
1000 & $6.1 \times 10^{-6}$ & 52.9 & 1.2 & 45.47 \\
\specialrule{.2em}{0.2em}{0.2em}
\multicolumn{5}{c}{\textbf{EHT array configuration}} \\
\specialrule{.05em}{0.2em}{0.2em} 
1000 & $7.6 \times 10^{-4}$ & 32.6 & 40.7 & - \\
\multicolumn{5}{c}{\textbf{ALMA array configuration}} \\
\specialrule{.05em}{0.2em}{0.2em} 
1000 & $4.6 \times 10^{-6}$ & 53.7 & 0.8 & -\\

\specialrule{.3em}{0.2em}{0.2em} 
\end{tabular}
\caption{MSE, PSNR, SRE, and the sampling time $t_\mathrm{sampling}$ as a function of number of sampling steps $K$. For every  $K$ the batch size is 128, thus the $t_\mathrm{sampling}$ represents the time to generate 128 reconstructions. We note that $99\%$ of the pixels in the ground truth image have very small flux values, below 5\% of the maximum value.}
\label{tab:mse_calibration_vs_steps}
\end{table}

We  evaluate the image reconstruction results for different numbers of sampling steps of the DDRM algorithm. Here, the number of sampling steps refers to the number of discrete iterations used during the reverse procedure. Sampling with $K$ timesteps where $K < T$ means selecting $K$ timesteps of our complete diffusion schedule $\beta$. Sampling with a larger number of steps should yield better results at the cost of more time, while using fewer timesteps would trade quality for sampling time.
We also evaluate the time to process the sampling steps, calculated using one NVIDIA GH200 GPU on the CSCS Alps infrastructure.

The results are shown in Table~\ref{tab:mse_calibration_vs_steps}.   All tests are run with a batch size of 128, thus creating 128 DDRM reconstructions per test. We find that MSE improves with higher values of $K$,  whereas SRE does not depend as strongly on the number of sampling steps. The best MSE of $6.1 \times 10^{-6}$ is achieved with the maximum sampling steps $T=1000$. However, an MSE of $3.2 \times 10^{-4}$ is achieved with only 10 sampling steps, which can run in 0.44s, demonstrating the efficiency of DDRM.

However, we find that when reducing the number of sample steps,  DDRM is less likely to recover faint image components. In the subsequent results, all DDRM reconstructions are run for 1000 sampling steps.

\subsection{Uncertainty estimation}

DDPMs are often used to quantify and model uncertainty
in various applications because of their inherent ability to learn the entire data distribution and therefore generate diverse, realistic samples that capture a range of possible outcomes. 
We evaluate how well the DDPM prior can accurately represent the reconstruction error through the SRE, which measures the reconstruction error relative to the variation in the reconstructions. A well-calibrated network should have $\mathrm{SRE}\leq1$.

We show the standard deviation of 128 reconstructions and the SRE in the two leftmost columns of Figure~\ref{fig:ddrmreco}. We find that overall the DDRM is overconfident, with values of up to $\mathrm{SRE}=8$ where the DDRM image reconstruction has failed to recover low-flux components of the target image. 
This result is not surprising, as DDPMs are by default uncalibrated~\citep{pang2023on}, ie variations in model outputs are smaller than the residuals.   The range of DDRM predictions does not  accurately represent the reconstruction error, and therefore cannot be used as an image uncertainty.
Section~\ref{sec:conclusion} presents a discussion of possible improvements and calibration methods.

  \begin{figure*}
\centering
  \includegraphics[width=.99\linewidth]{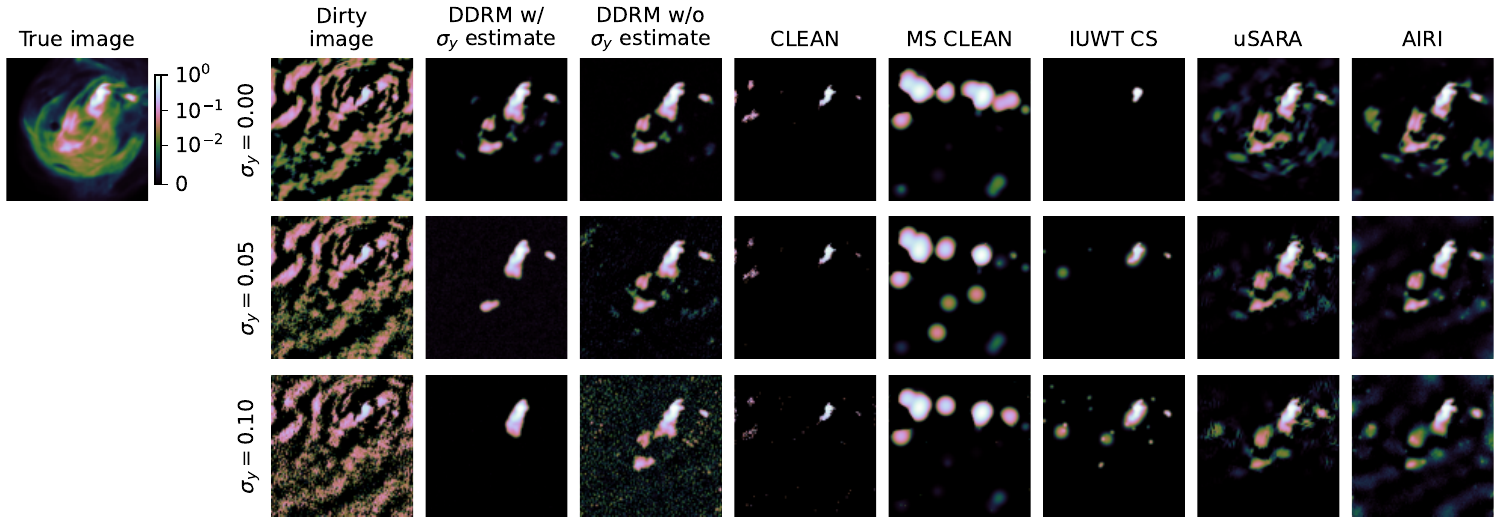}
  \caption{ Reconstruction results for a $150\times150$ pixel cutout of 3c353, an out-of-domain radio galaxy, imaged from a simulated VLBA observation, variable observation  noise added from $\sigma_y = 0$ to $0.1$. Columns from left to right are: the true image $\mathbf{x}$, the dirty image $\mathbf{x}_D$, the DDRM restored image using the true value of $\sigma_y$, the DDRM restored image using $\sigma_y = 0$, and then the CLEAN, multi-scale CLEAN, IUWT compressed sensing, AIRI, and uSARA reconstructions. The true and dirty images are scaled to the range [0,1], and the flux does not correspond to physical units. All log scale colors revert to linear scale in the range [0,0.01].
  }
  \label{fig:3c353}
\end{figure*}

\subsection{Comparison to other deconvolution algorithms}

We also compare the DDRM restoration results {  to five other deconvolution algorithms.

We run ``generic'' CLEAN, multi-scale (MS) CLEAN, and isotropic undecimated wavelet transform (IUWT) compressed sensing~\citep{2015A&A...576A...7D} as implemented in the Radler: Radio Astronomical Deconvolution Library\footnote{\href{https://radler.readthedocs.io/en/latest/index.html}{https://radler.readthedocs.io/en/latest/index.html}}, which is a module of WSCLEAN~\citep{offringa-wsclean-2014,offringa-wsclean-2017,vandertol-2018}. We run these Radler reconstructions for 100,000 iterations with a major loop gain of 1.0 and automatic thresholding with $\sigma =2$.}

 { 
We also compare DDRM against the AIRI and uSARA algorithms as implemented in the Small-scale-RI-imaging library\footnote{\href{https://github.com/basp-group/Small-scale-RI-imaging}{https://github.com/basp-group/Small-scale-RI-imaging}}. We define a new \texttt{MeasOp
} class that uses the same transformation operator as our DDRM sampling. \cite{2023MNRAS.518..604T} recommend using a heuristic of $\sigma_y/\sqrt{2L}$, where $L$ is the spectral norm of $\mathrm{Re} \{ \mathbf H^\dagger \mathbf H \}$ and we have $L=1$ for our transformation operator. In practice we find that $(0.001 + \sigma_y)/\sqrt{2L}$ yields the best image reconstruction quality. Otherwise we run both algorithms using the default settings.}

Image reconstruction results for these { five} algorithms and DDRM reconstruction are shown in Figure~\ref{fig:all} for a radio galaxy in the test dataset, with no noise added to the simulated observation. Unlike the CLEAN algorithms, the DDRM reconstruction does not contain any residual dirty beam artifacts. By eye, the DDRM reconstruction looks comparable to the  IUWT, uSARA, and AIRI reconstructions, but with additional noise which arises from the sampling process.

\begin{figure}
\centering
  \includegraphics[width=.99\linewidth]{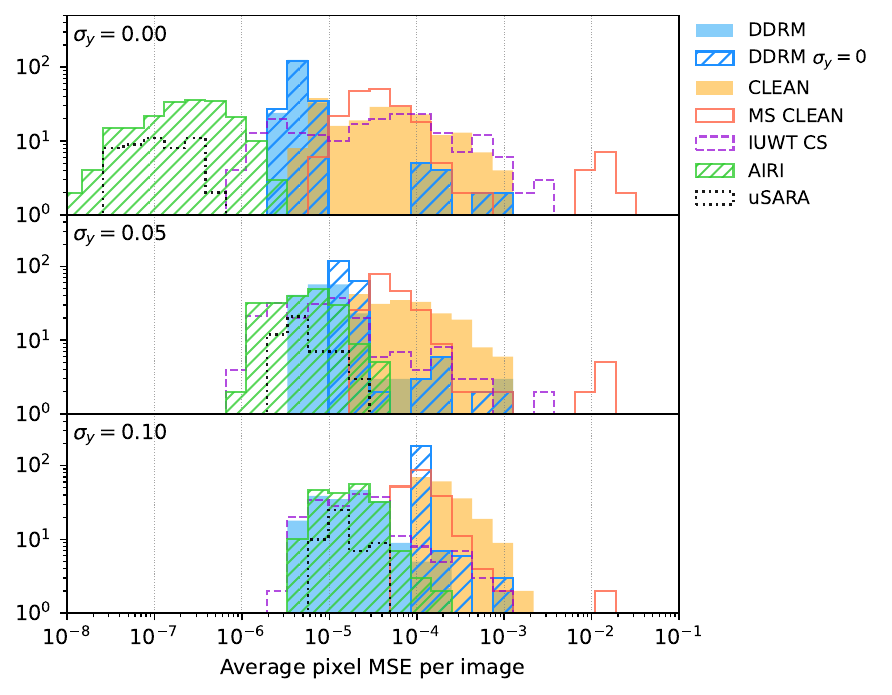}
  \caption{ Average MSE per image for 200 reconstructed images of the test data set. Reconstruction is done from a simulated VLBA observation  with  noise from $\sigma_y = 0$ to $0.1$. The DDRM restored image uses the true value of $\sigma_y$ (solid blue), or uses $\sigma_y = 0$ regardless of the true noise level (diagonal hashed blue). { The uSARA results are only calculated on the first 50 images of the test set due to the computational cost.}
  }
  \label{fig:recostats_avg}
\end{figure}

We also evaluated the reconstruction quality when including noise in the simulated observations. The  average MSE across all images for the different reconstruction methods are shown in Figure~\ref{fig:recostats_avg}.
We find that for all tests the DDRM image MSE is at least one order of magnitude smaller than CLEAN and MS CLEAN, even in the case of extremely noisy images with $\sigma_y = 0.1$ (noise level 10\% of maximum pixel value).
While uSARA and AIRI offer the best reconstruction quality at $\sigma_y = 0$, the reconstruction quality is similar to DDRM and IUWT CS as the noise level increases. We note that on an NVIDIA GH200 GPU, uSARA takes ~1-2 minutes to reconstruct an image, AIRI takes 15-20s, and DDRM takes 2-3s when reconstructing with $K=1000$ sampling steps and a batch size of 1.
Overall DDRM offers excellent reconstruction quality across a wide range of noise levels.

\subsection{Incorrect noise and out-of-domain tests}
\begin{figure}
\centering
  \includegraphics[width=.99\linewidth]{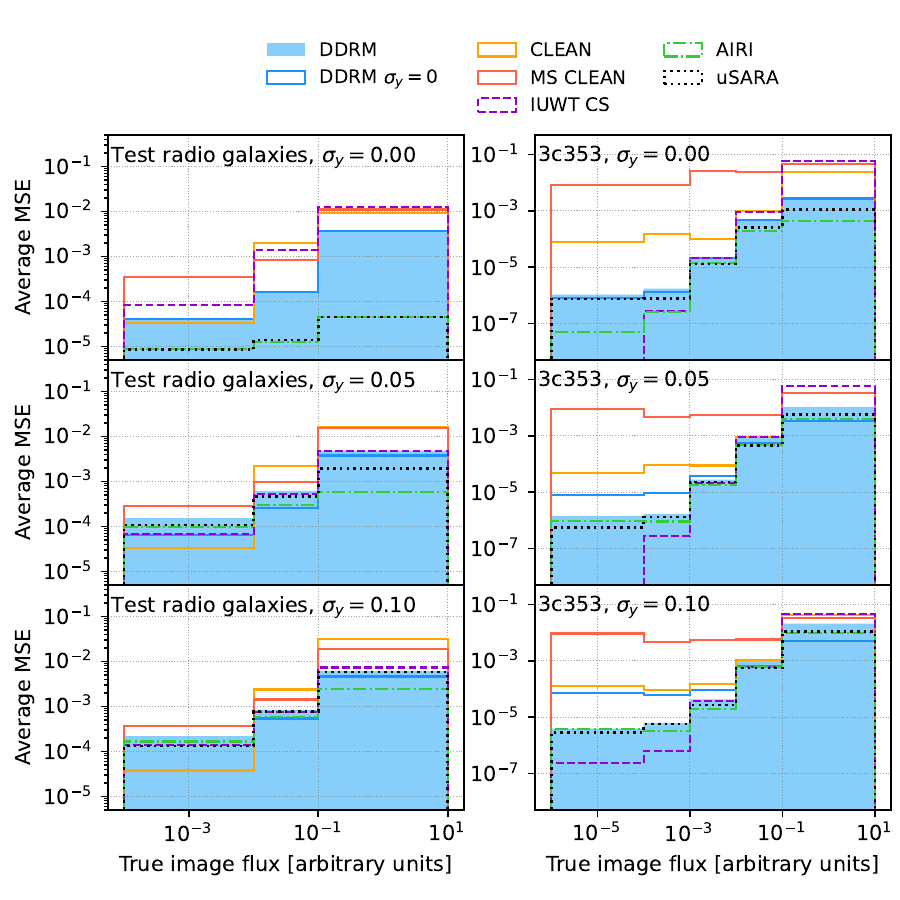}
  \caption{ MSE as a function of true image pixel brightness for (left) 200 images of the test data set and (right) several different $150\times150$ pixel cutouts of 3c353. Reconstruction is done from a simulated VLBA observation  with  noise from $\sigma_y = 0$ to $0.1$. The DDRM restored image uses the true value of $\sigma_y$ (solid blue), or uses $\sigma_y = 0$ regardless of the true noise level (blue line). { The uSARA results are only calculated on the first 50 images of the test set due to the computational cost.}
  }
  \label{fig:recostats}
\end{figure}

{ 
We also explore running the DDRM sampling with an incorrect estimate of the noise, always using $\sigma_y = 0$ regardless of the correct observation noise. We note that setting $\sigma_y = 0$ means that the network will inpaint missing visibilities but will not denoise  existing visibilities.
Figure~\ref{fig:recostats_avg} shows that even when  $\sigma_y = 0.1$ the reconstruction quality is better on average than CLEAN and MS CLEAN, demonstrating the robustness of the DDRM reconstruction.

We also test reconstruction on several  $150 \times 150 $ pixel cutouts of the radio galaxy 3c353 which was used as an imaging benchmark in~\cite{2023MNRAS.518..604T}. This image has a much larger dynamic range than our training set, exhibits compact morphology and extended diffuse structures, and challenges our learned DDPM prior. The radio galaxy and the corresponding reconstruction results are shown in Figure~\ref{fig:3c353}. Columns 2 and 3 show the reconstruction results when running DDRM with the true vaue of $\sigma_y$ vs always using $\sigma_y=0$. When DDRM sampling is run without the measurement noise estimate, it is better able to reconstruct the diffuse components of the true image, though the image overall is noisier.  The $\sigma_y$ parameter is used to balance sampling with the prior vs the measurement, so setting $\sigma_y=0$ means that the network will only use its low-dynamic range prior when inpainting missing visibilities. By eye it seems that DDRM with $\sigma_y=0$ is able to reconstruct the low-flux diffuse components of the image as well as uSARA or AIRI, even as noise increases.

The reconstruction MSE vs true image pixel intensity is shown in Figure~\ref{fig:recostats} for our test dataset and 3c353. For both datasets we see that DDRM consistently outperforms the CLEAN and MS CLEAN image reconstructions. DDRM outperforms IUWT CS in brighter regions of the image, though IUWT CS is better at low flux values likely because of its sparsity prior, whereas DDRM images contain some residual noise from the sampling process.
When $\sigma_y=0$ AIRI and uSARA outperform DDRM, but the performance of the three algorithms is very similar as noise increases. 
}



\section{Discussion \& Conclusions}

\label{sec:conclusion}

In this work, we present a new approach to radio interferometric image reconstruction using DDRMs. {  Our approach naturally incorporates the instrument PSF and noise.} { Unlike sparsity-based deconvolution algorithms, DDRM uses a data-driven prior trained on representative data. Unlike other machine learning approaches, DDRM does not need to be trained on the specific array layout.}

We find that the DDRM algorithm can propose a range of plausible restorations for the sub-sampled images typically recovered in radio interferometry. The method is less dependent than the CLEAN algorithm on user input,  and we also find that images are reconstructed with better MSE compared to CLEAN, MS CLEAN, or IUWT CS.  DDRM also offers a fast alterative to uSARA or AIRI for reconstructing noisy observations.

We note that we do not perform extensive hyperparameter optimization of the DDPM, and performance could be improved by searching over different network architectures or learning rate schedules.

We use a DDPM prior trained exclusively on VLA images. We do see that the performance of DDRM is best on the test VLA dataset, but we test on one out-of-domain galaxy 3c353 and find that the performance is still better than MS CLEAN and comparable to IUWT CS. We  note that the DDRM method can easily be used with a different DDPM such as the one developed by ~\cite{2024Vicanek} trained on the LOFAR Two-Metre Sky Survey \citep[LoTSS;][]{2022A&A...659A...1S} or the one developed by~\cite{2026arXiv260107485P} trained on MeerKAT data.  

Despite the {  high accuracy} of the DDRM reconstruction, there are several limitations:

\textbf{Image size:} An inherent constraint of the DDRM reconstruction method is the image size, limited to $150 \times 150$ pixel images, unlike CLEAN which can be run of images of any size. We are limited to this image size by the available training data of our DDPM. Simulated radio continuum survey maps such as~\citep{2025A&A...700A..18V} could provide larger image sizes to serve as a basis for training. 

 \textbf{Uncertainty:} 
 We did not find that the range of DDRM predictions accurately represented the reconstruction error, and therefore cannot be used as an image uncertainty. This could be improved by calibrating the DDPM~\citep{pang2023on}. Recent work by \citep{2023arXiv230203791T} has shown that conformal prediction can give finite-sample, distribution-free uncertainties for diffusion models. This method allows calibration of uncertainty thresholds on held-out data.
 
\textbf{Transformation operator:} DDRM relies on finding the SVD of the linear transformation operator $\mathbf H$. In this work we considered a relatively simple transformation consisting of a 2D Fourier transform and $(u,v)$ sampling. However, contemporary imaging techniques used in radio interferometry need to account for $A$-terms $W$-terms \citep{Bhatnagar_2008, Cornwell_2008_W_proj}, related to the direction-dependent antenna response and sky curvature, respectively. {  $A-$terms could be accounted for by including a matrix operator representing multiplication of the sky brightness by the primary beam $\mathbf A$:
\begin{equation}
    \mathbf y = \mathbf{ S \mathcal{F} A~x},
\end{equation}
extending Equation~\ref{eq:fwdmodel0}, though a new memory-efficient SVD would need to be defined for this operator. $W-$terms can be modeled using the non-uniform discrete Fourier transform \citep[NUDFT;][]{BagchiMitra1999NUDFT}, but we find that the SVD decomposition is intractably large for realistic observations due to the large number of visibility points.

Another way to use nonlinear transformation operators could be to use Diffusion Posterior Sampling~\citep{chung2023diffusion}, where update uses the full nonlinear degradation operator $H^T (y-Hx)$.

Any inaccuracies in the transformation operator arising from calibration mismatch would degrade the performance of DDRM. If the transformation operator is inaccurate, the DDRM reconstruction method can be weighted to use more contributions from the DDPM prior by tuning the $\eta_b$ hyperparameter, though this may degrade performance of out-of-domain sources.
}



\begin{acknowledgements}
      ET acknowledges financial support from the SNSF under the Starting Grant project Deep Waves (218396). RP acknowledges financial support from the SNSF under the Weave/Lead Agency project RadioClusters (214815). 
This work was supported by the Swiss National Supercomputing Centre (CSCS) under project ID {sk031}, done in partnership with the SKACH consortium through funding by SERI. 
\end{acknowledgements}

%
\bibliographystyle{aa} 
\bibliography{example} 
\begin{appendix}

\section{Uniform vs Natural UV Weights}

  \begin{figure}
\centering
  \includegraphics[width=.99\linewidth]{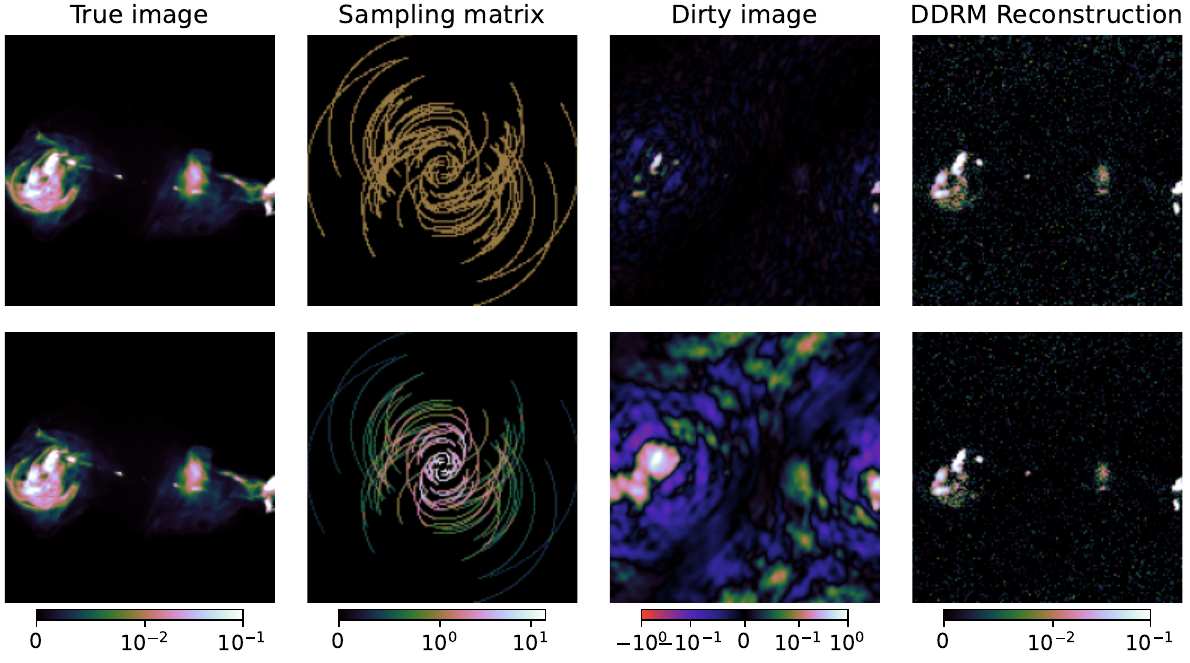}
  \includegraphics[width=.99\linewidth]{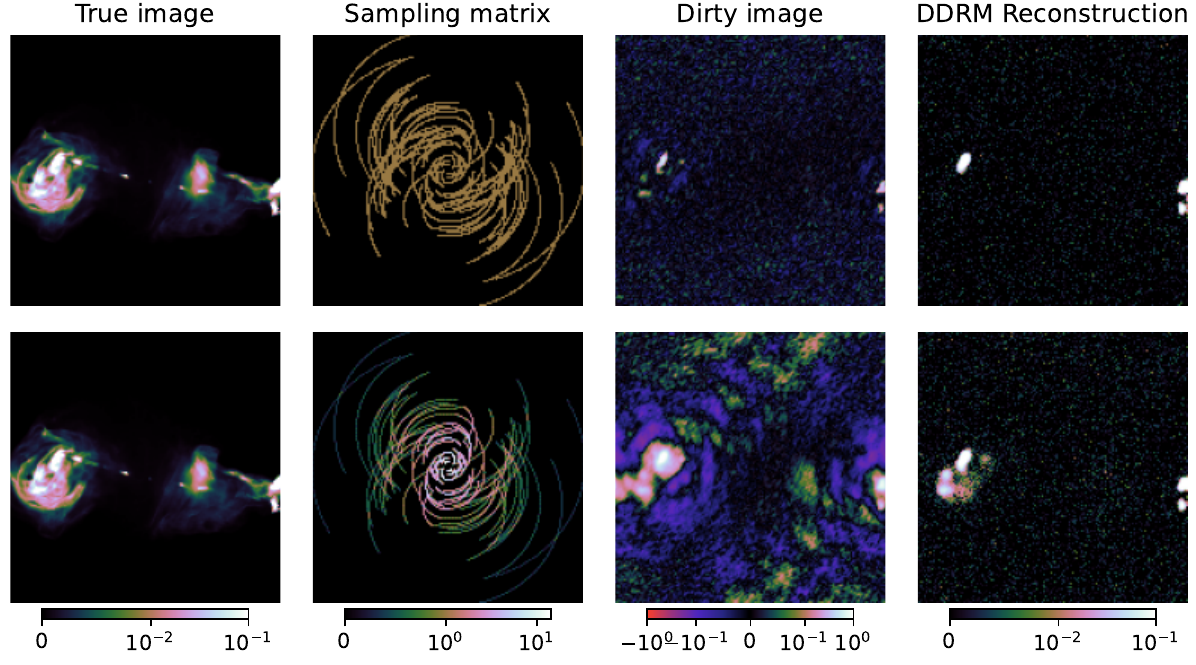}
  \caption{ Reconstructing a $150\times150$ pixel image of 3c353 using uniform (all uv bins get the same weight) and natural (uv bins are weighted according to the number of uv samples) weighting, with $\sigma_y=0$ (top) and $\sigma_y=0.05$ (bottom). 
  }
  \label{fig:recoweight}
\end{figure}
\label{sec:app:weights}
{ 
The transformation operator considered in throughout the paper corresponds to \emph{uniform}  weighting, giving all uv bins the same weight. Different weight schemes are easily implemented with DDRM, and corresponds to modifying the singular values of the operator. We also test \emph{natural}  weighting, giving each uv bin a weight proportional to the number of uv samples in the bin. When doing image reconstruction, DDRM will use the observation constraint more strongly in portions of visibility space that have a large weight compared to the observation noise, as discussed in Section~\ref{sec:ddrm}. The results are shown in Figure~\ref{fig:recoweight} {  for DDRM sampling using $\eta = 0.85$ and $\eta_b = 1.0$}. When the observation has no noise (top two rows), the reconstructions are equivalent. When the observation is noisy, DDRM trusts the higher-weight central baselines more, resulting in better reconstruction of large-scale diffuse emission.
}

\section{Additional out-of-domain image reconstruction tests}
\label{sec:app:ood}

\begin{figure}
\centering
  \includegraphics[width=.99\linewidth]{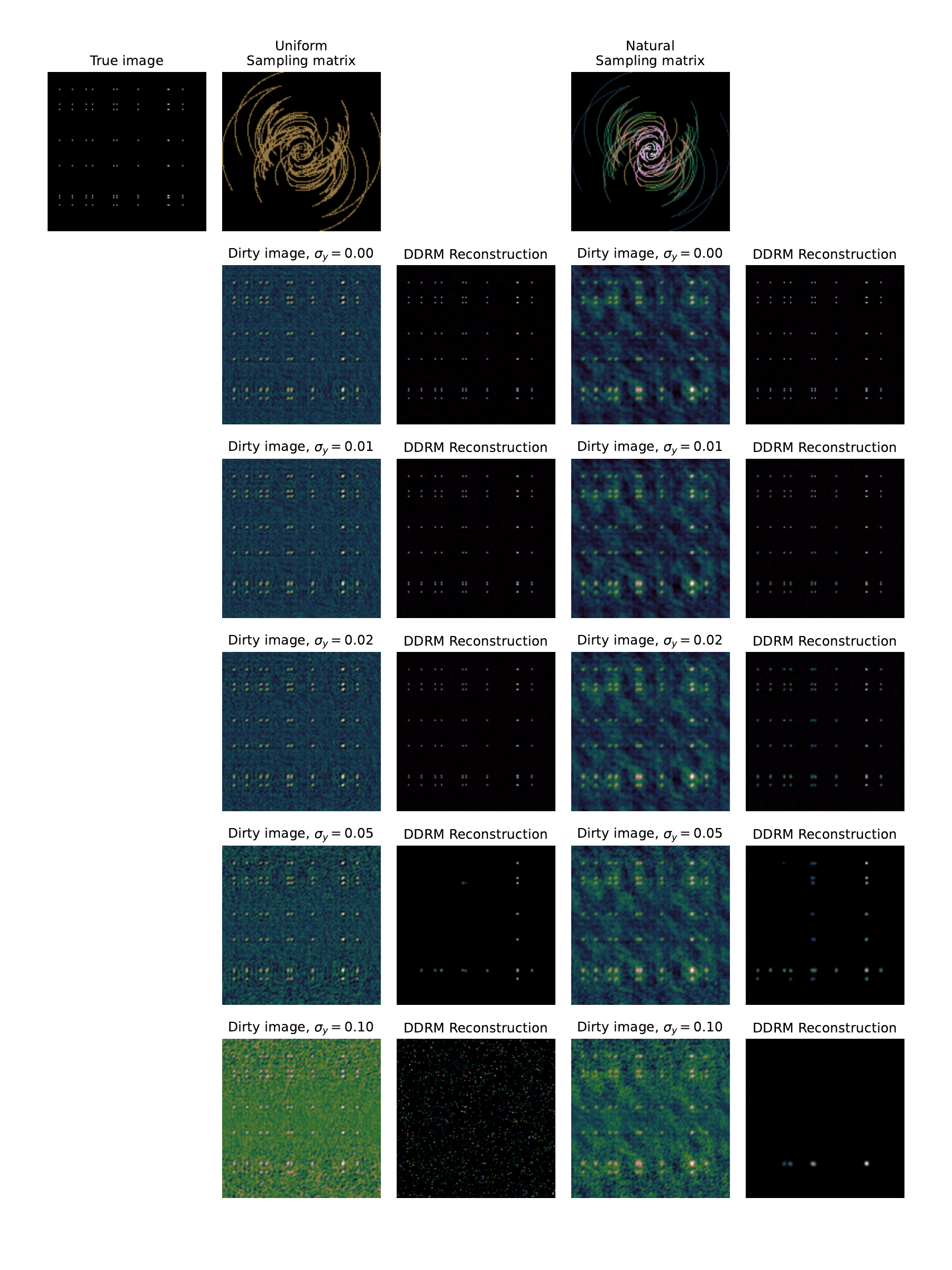}
  \caption{ Reconstructing a $150\times150$ pixel image of a random grid of single-pixel dots using uniform (all uv bins get the same weight) and natural (uv bins are weighted according to the number of uv samples) weighting, going from observation noise of $\sigma_y=0$ (top) and $\sigma_y=0.1$ (bottom). 
  }
  \label{fig:dots}
\end{figure}

\begin{figure}
\centering
  \includegraphics[width=.99\linewidth]{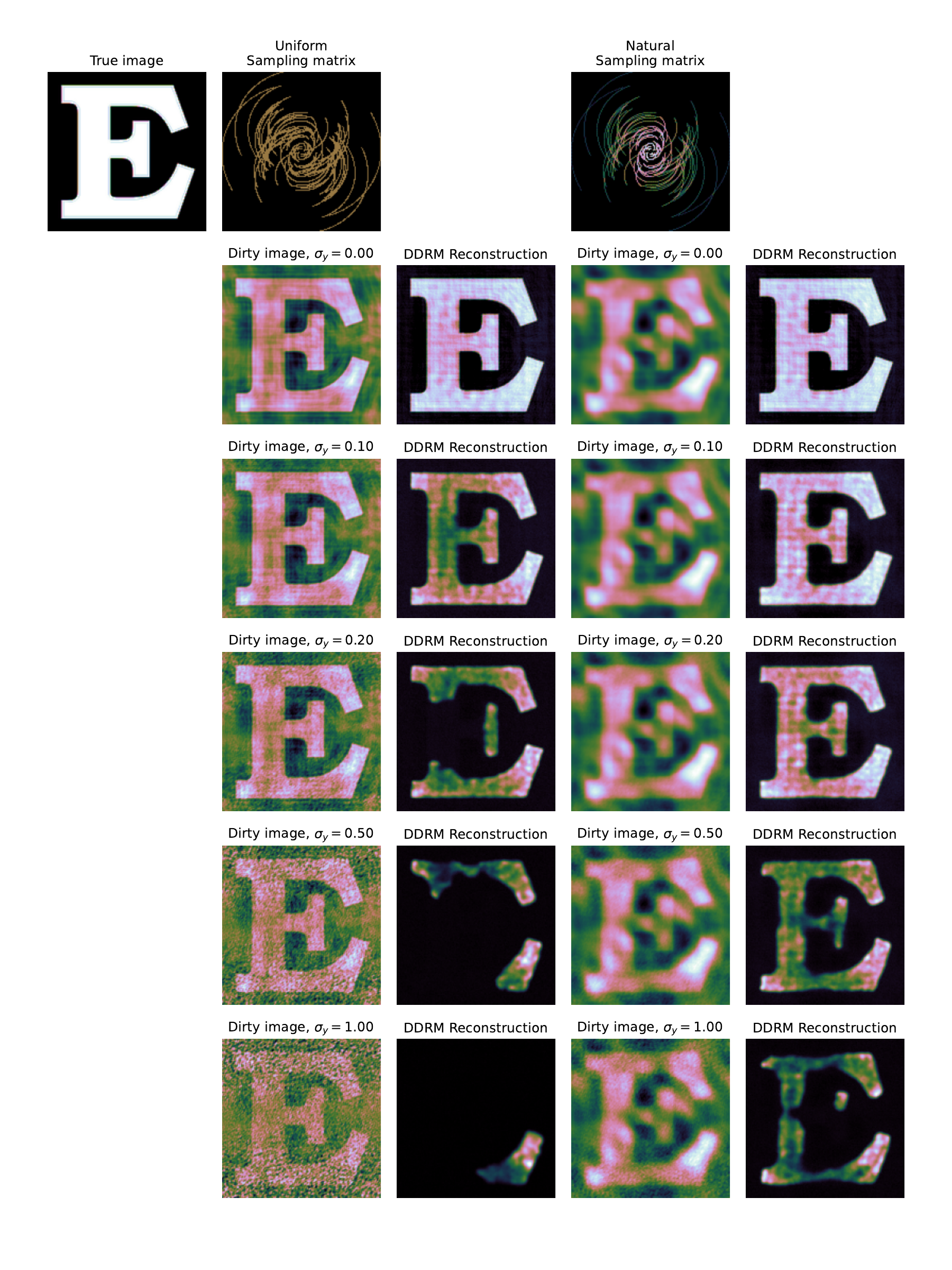}
  \caption{ Reconstructing a $150\times150$ pixel image  using uniform (all uv bins get the same weight) and natural (uv bins are weighted according to the number of uv samples) weighting, going from observation noise of $\sigma_y=0$ (top) and $\sigma_y=1.0$ (bottom). 
  }
  \label{fig:letterE}
\end{figure}

We test the DDRM reconstruction performance on two additional out-of-domain images, { a randomly-spaced grid of point sources (shown in Figure~\ref{fig:dots}) and the letter `E' (shown in Figure~\ref{fig:letterE}). These two images demonstrate the tradeoff between the prior and the observation in DDRM sampling, as well as the applicability of the prior to out-of-domain sources.} With no observation noise $\sigma_y = 0$, DDRM only inpaints the missing visiblity data using the prior from the DDPM. As $\sigma_y$ increases, the DDRM denoises the visibilities using the prior.
{  For the uniform weighting, all visibilities are given the same relative $\sigma_y$, so all observations are weighted qually with respect to the prior in Equation~\ref{eq:sampleuncertain}. With the natural weighting, the central visibilities have a higher weight and therefore a smaller relative $
\sigma_y$, so these visibilities are weighted more with respect to the prior during sampling. For these tests we ran DDRM sampling with $\eta = 0.85$ and $\eta_b = 1.0$.

Although our training dataset did not contain any point sources, the network is able to accurately reconstruct the grid of points in Figure~\ref{fig:dots}, likely because the ground truth is sparse. As the observation noise increases to $\sigma_y \geq 0.05$, DDRM samples more heavily with the prior, and some of the point sources are not reconstructed.

The letter `E' image in Figure~\ref{fig:letterE} is is unlike anything in the training set, with all pixels set to 0 except for a continuous region where all pixels are set to 1. We see that the DDRM reconstruction does not reconstruct the letter with perfectly uniform flux.
}

{



\section{Hyperparameter Optimization}
\label{sec:eta_tuning}

The choice of hyperparameter $\eta$ affects the reconstruction error as well as the SRE, as shown in Table~\ref{tab:eta_tuning}. 
We chose $\eta = 0.65$ for an optimal tradeoff  between PSNR($= 52.87)$ and SRE($1.15)$ for $K=1000$ sampling steps. We find that $\eta_b = 1.0$ always yields the best reconstruction results.

\begin{table*}
\centering
\caption{PSNR and SRE for varying $\eta$ across different numbers of sampling steps $K$. }
\label{tab:eta_tuning}
\centering
\begin{minipage}{0.18\textwidth}
\centering
\textbf{$K = 1000$}\\[4pt]
\begin{tabular}{cccc}
\hline
$\eta$ & $\eta_b$ & PSNR & SRE \\
\hline
0.40 & 1.00 & 54.08 & 0.70 \\
0.50 & 1.00 & 55.20 & 0.22 \\
0.60 & 1.00 & 54.33 & 0.57 \\
0.65 & 1.00 & 52.87 & 1.15 \\
0.70 & 1.00 & 52.64 & 1.61 \\
0.80 & 1.00 & 51.00 & 3.25 \\
0.90 & 1.00 & 49.66 & 5.59 \\
1.00 & 1.00 & 48.65 & 12.35 \\
\hline
\end{tabular}
\end{minipage}
\hspace{5em}
\begin{minipage}{0.18\textwidth}
\centering
\textbf{$K = 500$}\\[4pt]
\begin{tabular}{cccc}
\hline
$\eta$ & $\eta_b$ & PSNR & SRE \\
\hline
0.40 & 1.00 & 51.48 & 2.30 \\
0.50 & 1.00 & 53.02 & 1.13 \\
0.60 & 1.00 & 54.38 & 0.42 \\
0.65 & 1.00 & 53.72 & 0.39 \\
0.70 & 1.00 & 54.82 & 0.23 \\
0.80 & 1.00 & 54.09 & 0.61 \\
0.90 & 1.00 & 52.74 & 1.68 \\
1.00 & 1.00 & 51.18 & 6.56 \\
\hline
\end{tabular}
\end{minipage}
\hspace{5em}
\begin{minipage}{0.18\textwidth}
\centering
\textbf{$K  = 100$}\\[4pt]
\begin{tabular}{cccc}
\hline
$\eta$ & $\eta_b$ & PSNR & SRE \\
\hline
0.40 & 1.00 & 48.39 & 4.72 \\
0.50 & 1.00 & 48.65 & 4.24 \\
0.60 & 1.00 & 48.97 & 3.62 \\
0.65 & 1.00 & 47.50 & 4.26 \\
0.70 & 1.00 & 49.48 & 2.99 \\
0.80 & 1.00 & 50.32 & 2.45 \\
0.90 & 1.00 & 51.68 & 2.13 \\
1.00 & 1.00 & 55.15 & 1.70 \\
\hline
\end{tabular}
\end{minipage}
\\
\vspace{1em}
\begin{minipage}{0.18\textwidth}
\centering 
\textbf{$K  = 50$}\\[4pt]
\begin{tabular}{cccc}
\hline
$\eta$ & $\eta_b$ & PSNR & SRE \\
\hline
0.40 & 1.00 & 46.47 & 4.84 \\
0.50 & 1.00 & 46.89 & 4.69 \\
0.60 & 1.00 & 47.25 & 4.90 \\
0.65 & 1.00 & 44.34 & 7.89 \\
0.70 & 1.00 & 47.62 & 5.74 \\
0.80 & 1.00 & 48.11 & 8.33 \\
0.90 & 1.00 & 48.80 & 17.01 \\
1.00 & 1.00 & 50.47 & 12.42 \\
\hline
\end{tabular}
\end{minipage}
\hspace{5em}
\begin{minipage}{0.18\textwidth}
\centering
\textbf{$K= 10$}\\[4pt]
\begin{tabular}{cccc}
\hline
$\eta$ & $\eta_b$ & PSNR & SRE \\
\hline
0.40 & 1.00 & 37.93 & 5.5 \\
0.50 & 1.00 & 38.69 & 3.5 \\
0.60 & 1.00 & 39.24 & 2.2 \\
0.65 & 1.00 & 36.96 & 1.8 \\
0.70 & 1.00 & 39.60 & 1.5 \\
0.80 & 1.00 & 39.73 & 1.2 \\
0.90 & 1.00 & 39.71 & 1.0 \\
1.00 & 1.00 & 39.36 & 0.8 \\
\hline
\end{tabular}
\end{minipage}
\end{table*}

}


\FloatBarrier 
\clearpage

\end{appendix}
\end{document}